\documentclass[letterpaper,twocolumn,english,amsmath,amssymb,noshowkeys,superscriptaddress,prd,nopacs,nofootinbib,nobibnotes]{revtex4-2}
\usepackage{mathptmx}

\usepackage[T1]{fontenc}
\usepackage[utf8]{inputenc}
\usepackage{xcolor}
\usepackage{babel}
\usepackage{float}
\usepackage{amsmath}
\usepackage{amssymb}
\usepackage{graphicx}
\usepackage[pdfusetitle,
 bookmarks=true,bookmarksnumbered=false,bookmarksopen=false,
 breaklinks=false,pdfborder={0 0 1},backref=false,colorlinks=true]
 {hyperref}
\hypersetup{
 colorlinks=true,linkcolor=red,filecolor=green,urlcolor=blue,citecolor=magenta }

\makeatletter

\usepackage{babel}
\usepackage{graphicx}
\usepackage[caption=false]{subfig}
\usepackage{bm}
\usepackage{lipsum}

\makeatother

\begin{document}
\title{Non-hermitian phase transitions on a generalized Ellis-Bronnikov wormhole
bridge}
\author{José A. S. Lourenço}
\affiliation{Departamento de Física, Universidade Regional do Cariri - 57072-270,
Juazeiro do Norte, Ceará, Brazil }
\email{arceniofisica@gmail.com}

\author{Ygor Pará}
\affiliation{Centro Interativo de Ciência e Tecnologia da Amazônia, Universidade
Federal do Pará, 66075-123, Belém-PA, Brazil }
\email{ygor.para@icen.ufpa.br}


\author{J. Furtado}
\affiliation{Universidade Federal do Cariri (UFCA), Centro de Ci\^{e}ncias e Tecnologia, Juazeiro do Norte, CE, 63048-080, Brazil}
\email{job.furtado@ufca.edu.br}

\selectlanguage{english}%
\begin{abstract}
In this paper, we investigate the  emergence of non-Hermitian phase transitions on a quantum wormhole surface. We consider a single fermion whose dynamics are governed by the Dirac equation confined to move on a quantum wormhole surface. The effects of the geometry are taken into account using the tetrad formalism and the spin connection. The Dirac equation gives rise to two coupled first-order differential equations for each spinor component. The eigenvalues and eigenfunctions for each spinor component are computed numerically, and the non-Hermitian phase transitions are investigated in terms of the geometric features of the wormhole and the magnitude of the imaginary component of the mass.
\end{abstract}
\maketitle

\section{Introduction\label{sec:Introduction}}

The investigation of quantum mechanics on curved surfaces unveils fascinating phenomena. Quantum particles, even when confined to curved geometries such as tori \citep{GomesSilva:2020fxo} or catenoids \citep{SILVA2020126458}, display wave-like properties, including interference and diffraction. The curvature of these surfaces introduces geometric influences that profoundly impact particle behavior, giving rise to remarkable effects like quantized energy levels and geometric phases \citep{dacosta1,dacosta2}. Additionally, surface curvature can generate curvature-induced forces acting on particles, modifying their motion and dynamics \citep{dacosta1,dacosta2}. Examining quantum mechanics in curved spaces provides valuable insights into both the foundations of quantum theory and the nature of geometry.

Two-dimensional nanostructures, such as graphene \citep{katsnelson_2012,Geim:2007aeb,CastroNeto:2007fxn} and phosphorene \citep{Phosphorene}, play a crucial role in low-energy physics due to their distinctive properties, which are strongly shaped by geometry \citep{dacosta1,dacosta2,COSTAFILHO2021114639,Vanderley}. These materials also function as analog models for systems in high-energy physics \citep{deSouza:2022ioq,Geovas,IORIO,IORIO20111334,Capozziello,CVETIC20122617,Behnam}. Among them, the torus is particularly noteworthy for its pronounced curvature effects \citep{GomesSilva:2020fxo,Ye_and_Job2022}, with carbon nanotori being employed in fields such as nanoelectronics, biosensing, and quantum computing \citep{biosensors,PhysRevLett.97.016601}.

In recent years, the catenoid geometry has attracted significant interest. As a minimal surface, the two-dimensional wormhole geometry is equivalent to a catenoid, which represents a possible solution for traversable wormholes \cite{dandoloffsaxenajansen}. In Ref.\cite{gonzalez,pincak}, a nanotube was proposed as a bridge connecting a bilayer graphene system. To ensure a smooth transition between the layers, Refs.\cite{dandoloff} and \cite{dandoloffsaxenajansen} suggested modeling the bilayer and the bridge using a single catenoid surface. This approach is feasible due to the curvature of the catenoid, which is concentrated around the bridge and asymptotically vanishes \cite{dandoloffsaxenajansen}. For a non-relativistic electron, the curvature of the surface induces a geometric potential in the Schr"odinger equation. The impact of geometry, along with external electric and magnetic fields, on the graphene catenoid bridge was investigated in \cite{euclides}, where a single electron is described by the Schr"odinger equation on the surface. Moreover, the effect of a position-dependent mass on an electron confined to a catenoid bridge was examined in \cite{job}, where an isotropic position-dependent mass was proposed as a function of both the Gaussian and mean curvatures.


\begin{figure*}[ht!]
    \centering
    \includegraphics[scale=0.33]{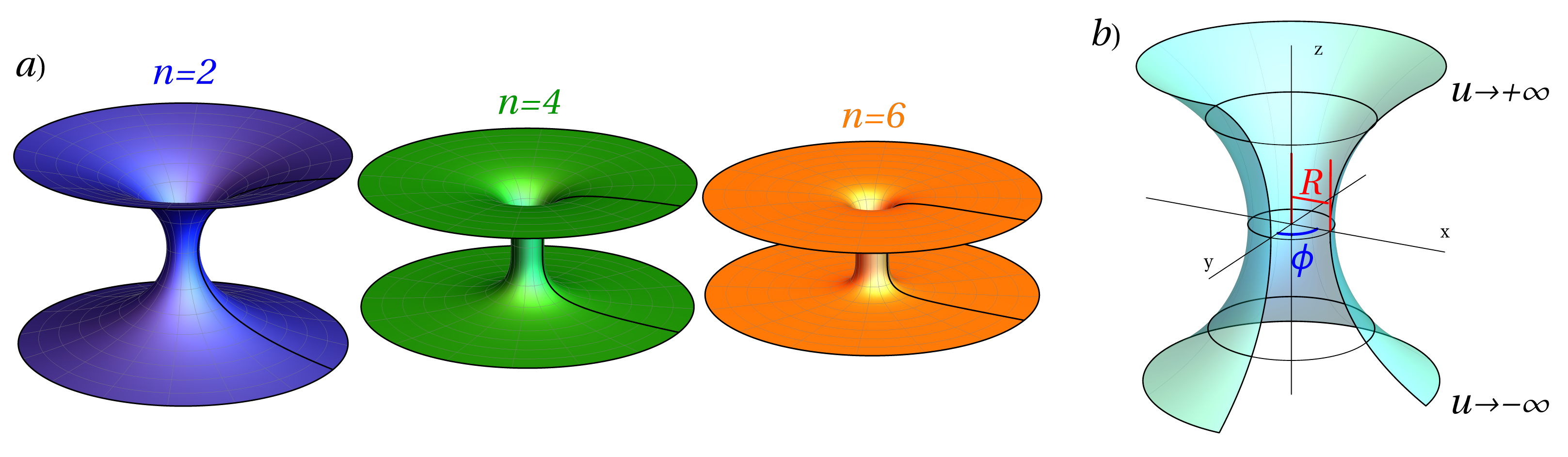}
    \caption{Wormhole geometry and the role of the deformation parameter $n$. (a) Embedding diagrams of the wormhole geometry for different values of the deformation parameter $n = 2, 4, 6$. As $n$ increases, the shape of the wormhole becomes flatter near the throat, with wider regions approaching a locally flat geometry away from the center. This illustrates the transition from a strongly curved to an almost cylindrical configuration as $n$ grows. 
    (b) Schematic representation of the wormhole profile in cylindrical coordinates $(u, \phi, z)$, indicating the throat radius $R$ and the angular coordinate $\phi$. The geometry connects two asymptotically flat regions as $u \to \pm \infty$.}
    \label{fig1}
\end{figure*}

Non-Hermitian systems have attracted significant attention across various fields of physics in recent years. Specifically, non-Hermitian systems that exhibit simultaneous parity $\mathcal{P}$ and time-reversal $\mathcal{T}$ symmetry can undergo $\mathcal{PT}$ phase transitions \citep{Bender}. This $\mathcal{PT}$-transition describes a shift in the system’s spectrum, where eigenvalues transition from real to complex or purely imaginary, and vice versa \citep{Bender2007}. These phase transitions are associated with critical points, known as exceptional points, which define the boundary between $\mathcal{PT}$-symmetric and $\mathcal{PT}$-broken phases \citep{Rotter,Heiss2012,Wei,PhysRevB.98.085126}. Various non-Hermitian phase transitions have been explored in different contexts, including mesoscopic systems with long-range interacting particles \citep{Lourenco22}, Floquet quasicrystal systems \citep{Zhou22}, coupled acoustic cavities with asymmetric losses \citep{Ding16}, anisotropic exciton-polariton pairs \citep{Chakrabarty23}, and non-Hermitian sensing techniques for controlling quasi-parametric amplifications \citep{Wu24}.

Non-Hermitian Hamiltonians naturally emerge in curved spaces, as observed in the quantum Beltrami surface \citep{FURTADO2023Electronic, Lambiase, IORIO2012334}, the generalized Ellis-Bronnikov graphene wormhole-like surface \citep{deSouza:2022ioq}, the torus \citep{GomesSilva:2020fxo}, and the catenoid \citep{SILVA2020126458,SILVA2021}, all of which exhibit $\mathcal{PT}$ symmetry in their Hamiltonians. Furthermore, non-Hermiticity can be introduced by an imaginary Dirac mass term \citep{Wang,Nori,Ygor2021}. Specifically, on the surface of the sphere \citep{Ygor2021}, there exists an infinite sequence of exceptional points (EP) that depend on the radius (curvature), resulting in non-Hermitian phase transitions.

In this paper, we investigate the  emergence of non-Hermitian phase transitions on a quantum wormhole surface. We consider a single fermion whose dynamics are governed by the Dirac equation confined to move on a quantum wormhole surface. The effects of the geometry are taken into account using the tetrad formalism and the spin connection. The Dirac equation gives rise to two coupled first-order differential equations for each spinor component. The eigenvalues and eigenfunctions for each spinor component are computed numerically, and the non-Hermitian phase transitions are investigated in terms of the geometric features of the wormhole and the magnitude of the imaginary component of the mass.

\section{Model\label{sec:Model}}

The dynamics of a massive fermion in a curved spacetime is described
by the Dirac equation for a two-component spinor $\ensuremath{\Psi(\mathbf{x})},$
given by 
\begin{equation}
\left(i\bar{\gamma}^{\mu}\nabla_{\mu}-M\right)\Psi\left(\mathbf{x}\right)=0,
\end{equation}
where $\bar{\gamma}^{\mu}=e_{A}^{\mu}\gamma^{A}$ are the gamma matrices
in curved space. We adopt the chiral representation $\ensuremath{\gamma^{0}=\sigma^{3}},$
$\gamma^{1}=-i\sigma^{2}$, and $\gamma^{2}=-i\sigma^{1}$. The vielbein
$e_{\mu}^{A}$ and its inverse $e_{A}^{\mu}$ satisfy the relation
$g_{\mu\nu}=\eta_{AB}e_{\mu}^{A}e_{\nu}^{B},$ where $\eta_{AB}=\text{diag}(-1,1,1)$
is the Minkowski metric. The spin connection is computed as $\omega_{AB\mu}=\eta_{\alpha\beta}e_{A}^{\alpha}\left(\partial_{\mu}e_{B}^{\beta}+\Gamma_{\mu\lambda}^{\beta}e_{B}^{\lambda}\right),$
where $\Gamma_{\mu\nu}^{\lambda}$ are the Christoffel symbols, given
by $\Gamma_{\mu\nu}^{\lambda}=(1/2)g^{\lambda\rho}(\partial_{\mu}g_{\rho\nu}+\partial_{\nu}g_{\rho\mu}-\partial_{\rho}g_{\mu\nu}).$

We analyze the dynamics of relativistic fermions constrained to the
surface of a Generalized Ellis-Bronnikov (GEB) wormhole. The geometry
is parametrized by the coordinates $(u,\phi)$, where $u\in(-\infty,\infty)$
represents the meridional coordinate, and $\phi\in[0,2\pi)$ is the
azimuthal angle. The line element in the equatorial slice $\theta=\pi/2$
is given by 
\begin{equation}
ds^{2}=-dt^{2}+du^{2}+f_{n}^{2}(u)d\phi^{2},
\end{equation}
where 
\begin{equation}
f_{n}(u)=\left(u^{n}+R^{n}\right)^{1/n}.
\end{equation}

Here, $R$ is the radius of the bridge, and $n$ is restricted to
even integers $(n=2,4,6,\dots)$ to preserve axial symmetry. The nonzero
components of the spin connection in this background are given by
$\omega_{12\phi}=-\omega_{21\phi}=-u^{n-1}f_{n}^{(1/n)-1},$
which leads to the Fock-Ivanenko coefficients $\Gamma_{\phi}=\frac{u^{n-1}\gamma^{1}\gamma^{2}}{4f_{n}^{1-1/n}}.$

\begin{figure}[b]
\begin{centering}
\includegraphics[width=1\columnwidth]{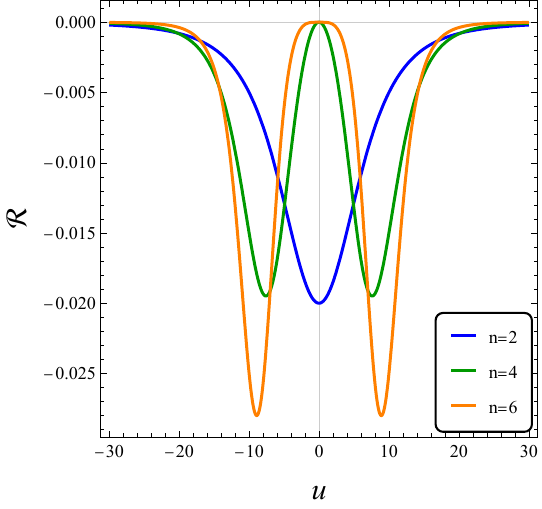}
\par\end{centering}
\caption{Curvature for the Generalized Ellis-Bronnikov wormhole with $R=10$ for three values
of the deforming parameter $n$, namely, $n=2$, $n=4$ and $n=6$.\label{fig:Curvature-for-the}}
\end{figure}

The Ricci scalar of the wormhole surface can be straightforwardly
computed, yielding 
\begin{equation}
\mathcal{R}=-2(n-1)R^{n}u^{n-2}f_{n}^{-2}.
\end{equation}
For $n=2$, the curvature remains negative everywhere and asymptotically
vanishes, as expected. However, for $n\neq2$, the surface deformation
gives rise to two regions of high curvature, separated by a cylindrical
region $(\ensuremath{\mathcal{R}=0})$ around $u=0$. The behavior
of the Ricci scalar as a function of $u$ is illustrated in Fig. \ref{fig:Curvature-for-the}. 

\begin{figure}[t]
\begin{center}
{\includegraphics[width=1\columnwidth]{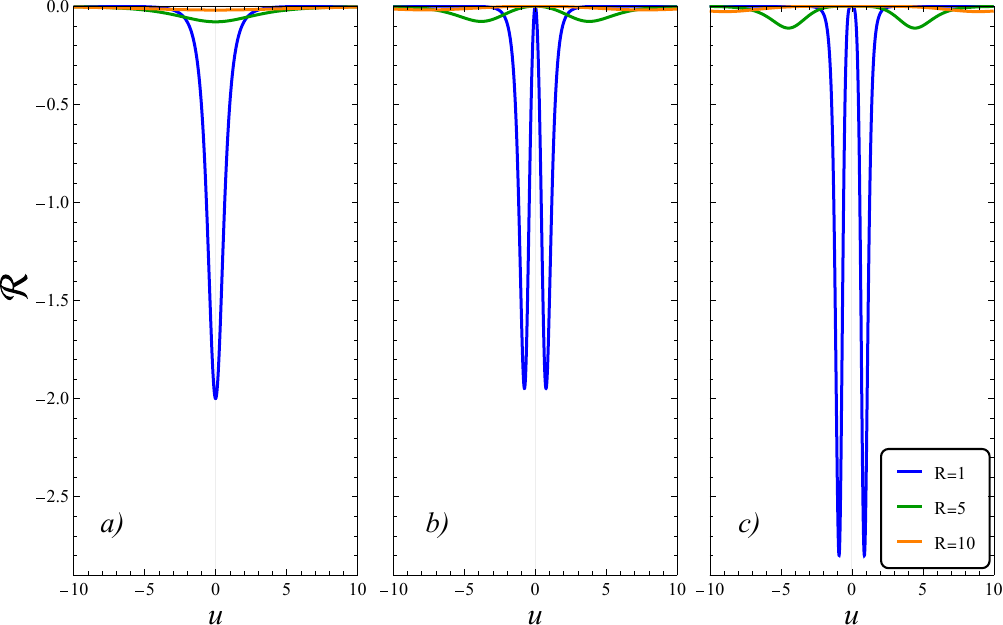}}
\end{center}
\caption{Effect of curvature for three values of the deforming parameter $n$, namely $n = 2$, $n = 4$, and $n = 6$, shown in panels (a), (b), and (c), respectively. In each case, the curves correspond to different curvature radii $R = 1$, $5$, and $10$, as indicated. The influence of curvature increases as the radius $R$ decreases.
    \label{fig:Curvature-for-the-1}}
\end{figure} 

To further investigate the impact of geometry on the system, we analyze the scalar curvature \(\mathcal{R}\) as a function of the radial coordinate \(u\) for different values of the deformation parameter \(n\) and curvature radii \(R\). As shown in Fig.~\ref{fig:Curvature-for-the-1}, panels (a), (b), and (c) correspond to \(n = 2\), \(4\), and \(6\), respectively. For each value of \(n\), we plot the curvature profiles for \(R = 1\), \(5\), and \(10\). The results reveal that, while the general structure of \(\mathcal{R}(u)\) remains centered around the wormhole throat, the amplitude and sharpness of the curvature peak increase significantly as \(R\) decreases. Additionally, higher values of \(n\) introduce more localized features near the throat, with the regions away from it becoming flatter, as expected from the geometric deformation. These behaviors highlight the interplay between the deformation parameter \(n\) and the curvature radius \(R\) in shaping the underlying geometry.

The Dirac equation can be separated into spatial and temporal components,
leading to a Schrödinger-like form $i\partial_{0}\Psi=H_{D}\Psi.$
We employ the \emph{ansatz} 
\begin{equation}
\Psi\left(u,\phi\right)=e^{-iEt-im\phi}\begin{pmatrix}\psi_{1}\left(u\right)\\
\psi_{2}\left(u\right)
\end{pmatrix},
\end{equation}
where $E$ is the energy eigenvalue and $m\in\mathbb{Z}$ is the azimuthal
quantum number. Consequently, the Dirac Hamiltonian takes the form
\begin{equation}
H_{D}=\begin{pmatrix}M & -i\mathcal{D}^{-}\\
-i\mathcal{D}^{+} & -M
\end{pmatrix},
\end{equation}
where the differential operators are defined as 
\begin{equation}
\mathcal{D}^{\pm}=\partial_{u}-\frac{u^{n-1}}{2f_{n}}\pm\frac{m}{\sqrt[n]{f_{n}}}.
\end{equation}

The study of open quantum systems offers valuable insights into the
dynamics of fermions in curved space. A widely used approach relies
on the Lindblad master equation, where the system’s evolution is governed
by an effective Hamiltonian that includes a non-Hermitian contribution,
responsible for dissipation and decoherence. A strikingly similar
mathematical structure emerges when introducing an imaginary mass
term in the Dirac equation, $M\to i\Gamma$. Analogous to the non-Hermitian
contribution in open quantum systems, this modification alters the
spectral properties of the system, leading to novel phase transitions.
This correspondence highlights a connection between non-Hermitian
relativistic quantum mechanics and dissipative dynamics, providing
a broader framework to explore emergent phenomena in both condensed
matter and high-energy physics.

To gain further insight into the effects of curvature and non-Hermitian dynamics, we now perform a numerical analysis of the Dirac equation in the GEB wormhole background. 
The energy spectrum is obtained by solving a second-order differential equation derived from the squared
Dirac Hamiltonian, $H_{D}^{2}\Psi=E^{2}\Psi$. 
This approach is motivated by the fact that the direct application of $H_{D}\Psi=E\Psi$ leads to a set of coupled first-order equations, making the second-order formulation more convenient for numerical computations.
In particular, we compute the energy eigenvalues by solving the following second-order differential equation 
\begin{equation}
-\partial_{u}^{2}\Psi+\frac{u^{n-1}}{f_{n}}\partial_{u}\Psi-\Upsilon_{n,m}\left(u,\Gamma\right)\Psi=E^{2}\Psi,
\end{equation}
where 
\begin{align}
\Upsilon_{n,m}\left(u,\Gamma\right) & =\frac{\left(n+\frac{1}{2}\right)u^{2n-2}}{2f_{n}^{2}}-\frac{\left(n-1\right)u^{n-2}}{2f_{n}}-\frac{\sigma_{z}mu^{n-1}}{f_{n}\sqrt[n]{f_{n}}}+\nonumber \\
 & -\frac{m^{2}}{\sqrt[n]{f_{n}}\sqrt[n]{f_{n}}}+\Gamma^{2}
\end{align}

The solutions for $E$ will reveal the interplay between geometry and non-Hermitian effects, which we discuss in detail in the next section.

\section{Results\label{sec:Results}}

\begin{figure*}[t]
\begin{centering}
\includegraphics[scale=0.35]{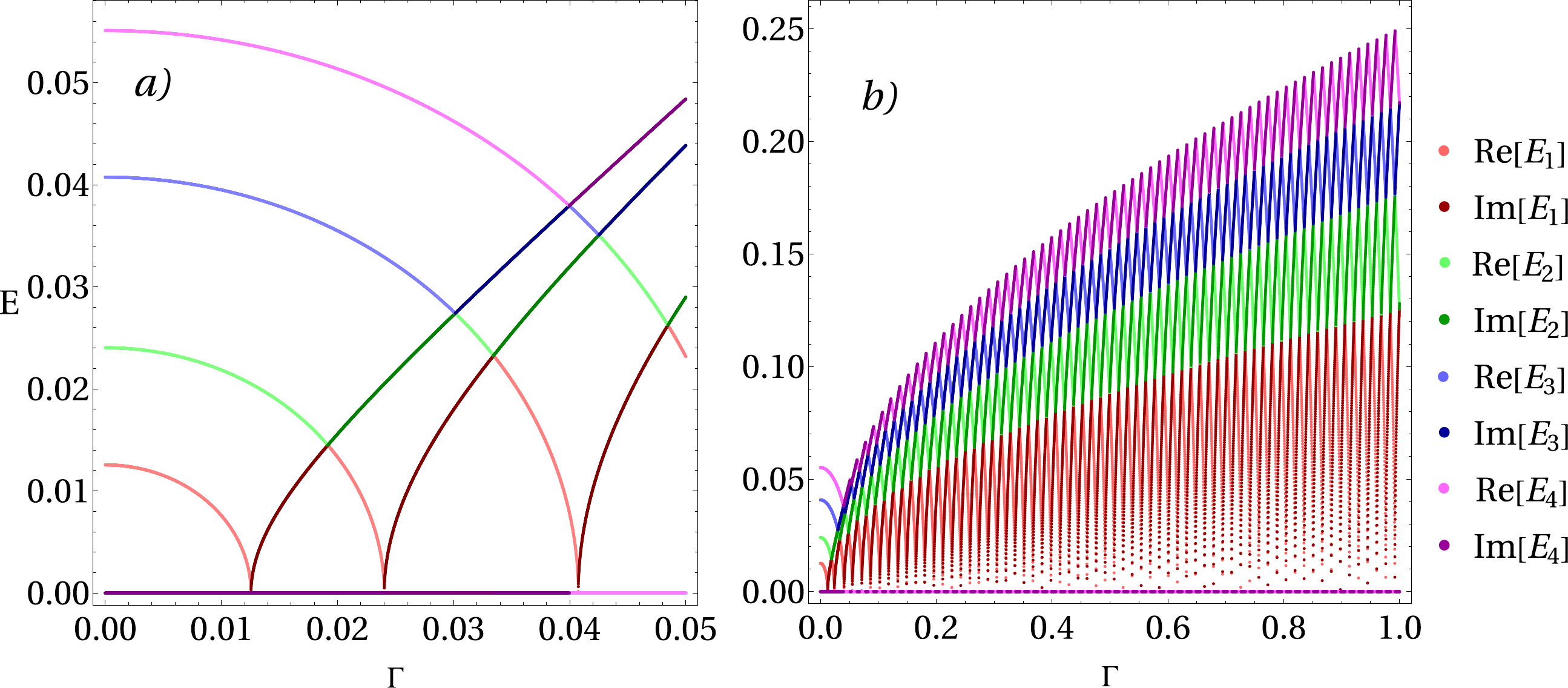}\end{centering}
\caption{First four positive eigenvalues for $R=1$ and $n=2$. Real (Re) and imaginary
(Im) components of the eigenvalues $E_{1},$ $E_{2},$ $E_{3},$ and
$E_{4},$ as functions of $\Gamma$. The confined oscillations of
the curves highlight the formation of exceptional points (EPs) at
intersections, where eigenvalues transition between real and complex
regimes. The reconnection of trajectories after each EP sustains the
system’s periodicity.\label{fig:First-four-eigenvalues}}
\end{figure*}

In this section, we present the numerical solutions of the differential
equation derived in the previous section and analyze the corresponding
eigenvalue spectrum. The obtained eigenvalues exhibit intriguing behavior,
particularly in their real components. As shown in Fig.~\ref{fig:First-four-eigenvalues}(a),
the red curve, which represents the ground state energy, demonstrates
a loss-like behavior as it approaches zero. This trend is a hallmark
of dissipation and aligns with the expectations for non-Hermitian
systems. Furthermore, an interesting feature emerges in the eigenvalue
$E_{2}$, which decays and eventually intersects with the $E_{1}$
curve, leading to the formation of a new exceptional point. A striking
aspect of this spectrum is how the curves interweave and merge into
one another forming a continuous and structured evolution of eigenvalues.
In particular, the $E_{4}$ curve (magenta) does not remain a distinct
entity but instead transitions through different eigenvalue branches
before reaching zero, indicating that it is composed of contributions
from other eigenvalues at different stages. Similarly, the other curves
follow distinct yet interconnected trajectories, with their imaginary
counterparts (darker blue and purple) exhibiting analogous trends.
The formation of exceptional points is clearly observed at the crossings
of these curves, where eigenvalues coalesce consistent with previous
studies on non-Hermitian phase transitions \cite{torusnh}.

\begin{figure}[b!]
\begin{centering}
\includegraphics[width=1\columnwidth]{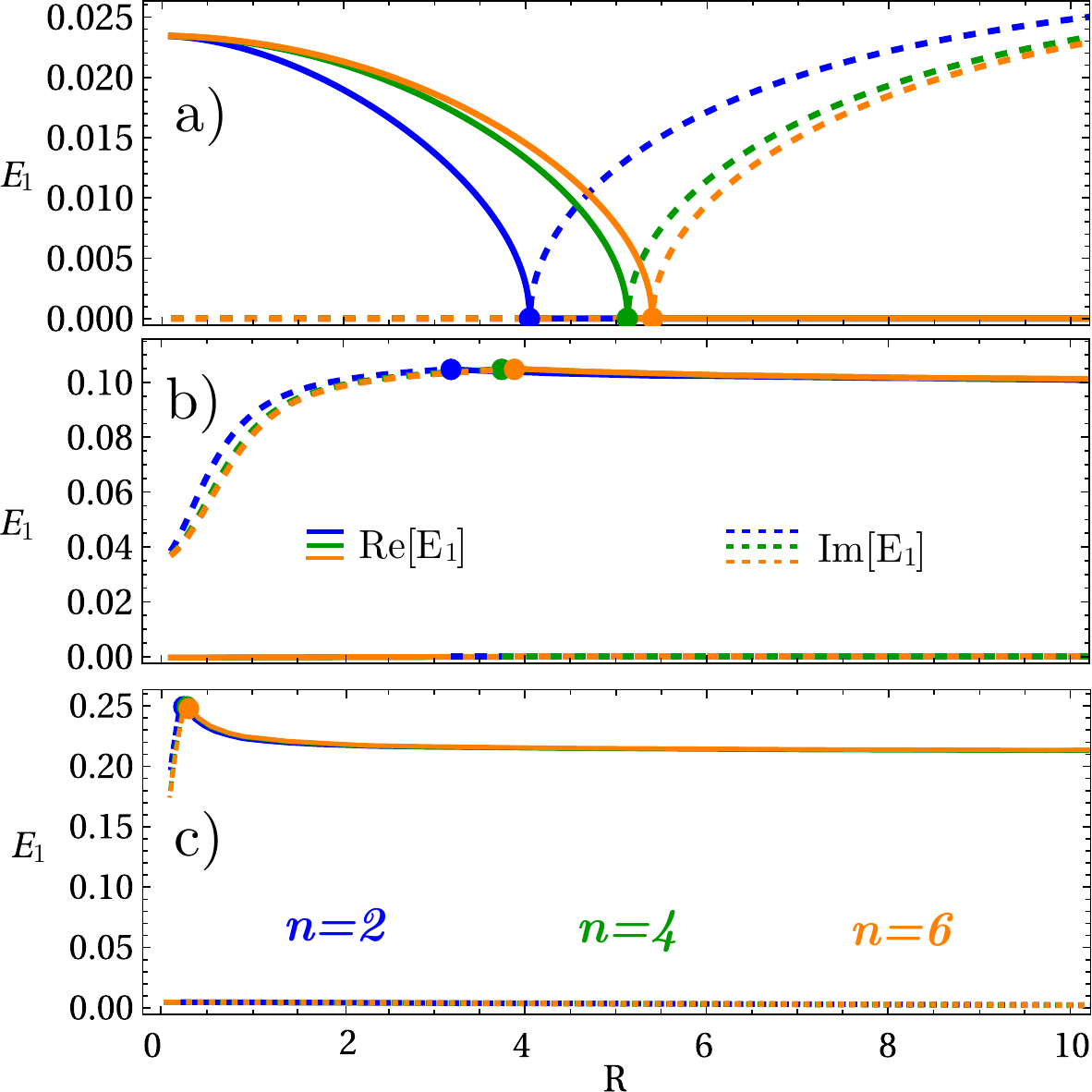}
\par\end{centering}
\caption{a) $\Gamma=0.1$, b) $\Gamma=0.7$ and c) $\Gamma=4$.\label{fig:E1xR}}
\end{figure}

The Fig.~\ref{fig:First-four-eigenvalues}(b) depicts the dynamics
of eigenvalues in a non-Hermitian system. Each eigenvalue oscillates
within a spatially confined region. The collision of two eigenvalues
generates an infinite sequence of exceptional points (EPs), which,
as previously noted, emerge from the intersection of their corresponding
trajectories. Initially, an eigenvalue exhibits purely real decay.
Upon reaching an exceptional point, it acquires complex components,
leading to imaginary growth. Subsequently, when intersecting the trajectory
of another eigenvalue, it transitions back to a real decay regime,
thereby completing the dynamical cycle.

\begin{figure*}[ht!]
    \centering
    \includegraphics[scale=0.5]{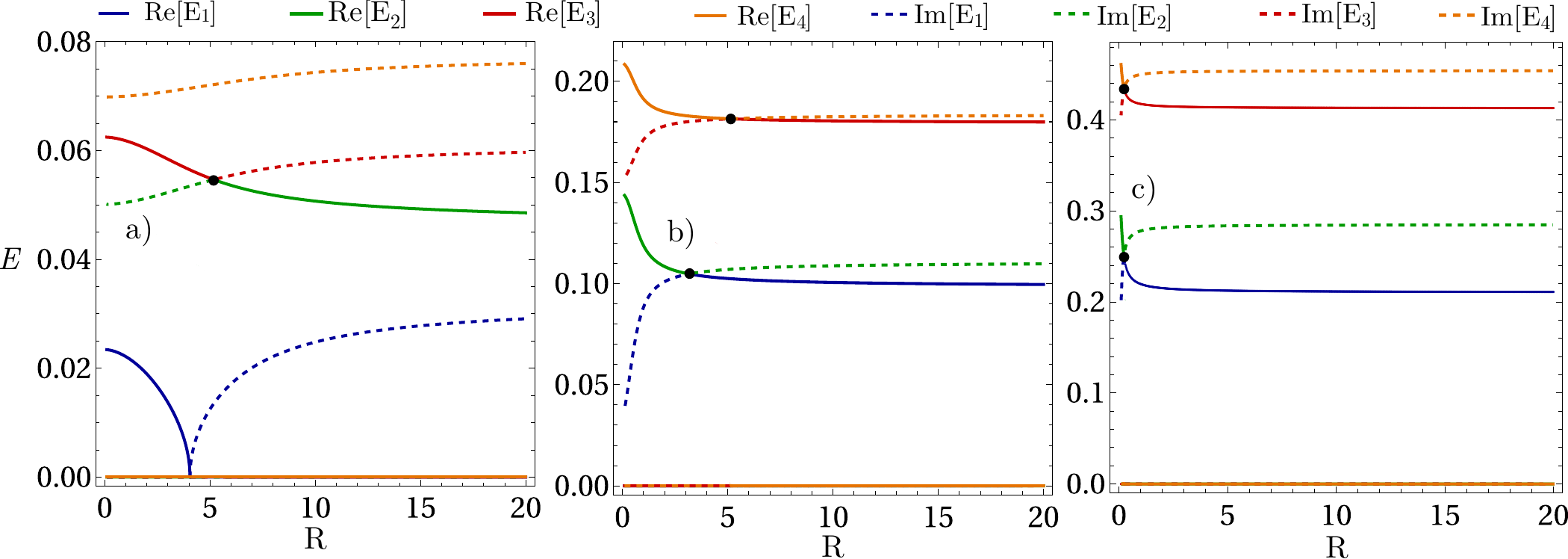}
    \caption{First four positive eigenvalues of $E$ with variation of $R$ for n = 2. a) $\Gamma=0.10$, b) $\Gamma=0.70$ and c) $\Gamma=4.$ }
    \label{fig:ExR}
\end{figure*}

In Fig. \ref{fig:ExR}, we show that changing the geometry can alter the behavior of the exceptional points. For simulations, we fix $m=1/2$, $n=2$, $R=\left[0.1, 20\right]$ and $u=\left[-100,100\right]$. This range of $u$ ensures that we can access the flat region of the wormhole \cite{deSouza:2022ioq}. We present the non-Hermitian phase transitions for the first four eigenvalues. The real part of the energy is represented by the solid line and the imaginary part by the dotted line. In Fig.~\ref{fig:ExR}~(a) the case of $\Gamma=0.10$ is observed, the eigenvalue $E_1$ (blue line) is initially real, then goes to zero at $R=4$ and becomes complex, this is the exceptional point of $E_1$. At the black point, the eigenvalues $E_2$ and $E_3$ share the same exceptional point (at $R= 5.17$) and $E_4$ admits only complex values. However, when we increase $\Gamma$ in Fig.~\ref{fig:ExR}~(b) to $\Gamma=0.7$ the eigenvalues $E_1$ and $E_2$ share the same exceptional point at $R=3.18$, and the eigenvalues $E_3$ and $E_4$ have the same exceptional point at $R=5.17$. In Fig.~\ref{fig:ExR}~(c) we use $\Gamma=4$.   

In Fig.~\ref{fig:E1xR}, we show the behavior of the exceptional points for the first eigenvalue $E_1$ when the radius $R$ is changed in the wormhole configurations of $n=2$ (blue line), $n=4$ (green line), and $n=6$ (orange line). The real part of $E_1$ is represented by the solid line and the imaginary part by the dotted line. In Fig.~\ref{fig:E1xR} we fix $\Gamma=0.1$ in (a), $\Gamma=0.7$ in (b) and $\Gamma=4$ in (c), and see that the exceptional points are displaced in all three cases when $n$ is changed. Note that increasing $\Gamma$ induces a smaller displacement of the exceptional points. Therefore, we observe that the eigenvalues have a similar behavior to the curvature parameter $\mathcal{R}$, dependent on the deformation parameter $n$, when $n$ increases the exceptional points tend to the same value (see Fig. \ref{fig:E1xR}), while the curvature parameter presents small changes in the shape of the curves (see Fig. \ref{fig:Curvature-for-the-1} (b) and Fig.\ref{fig:Curvature-for-the-1} (c)). Thus, for large values of $n$ the eigenvalues of the wormhole geometry present exceptional points at the same point.

\section{Experimental Feasibility and Physical Aspects\label{sec:Experimental-Feasibility-and}}

Let us now briefly examine the experimental feasibility of our model and the physical implications arising from the previously discussed results. To begin with, the fabrication of a graphene wormhole-like structures is an essential requirement for the proposed physical system. Recent advancements in lithographic techniques have already facilitated the development of wormhole-like arrays connecting graphene sheets~\cite{Lehr:14}, demonstrating the feasibility of constructing wormholes at the quantum scale. Single-walled graphene structures exhibit a linear dispersion relation and, therefore, their proper theoretical description necessitates the use of the Dirac equation. In contrast, multi-walled graphene structures possess a quadratic dispersion relation, so that the system's dynamics follows the Schrödinger equation. In our approach, we are considering single walled structures.

The subsequent challenge involves successfully confining a single electron within the system. While theoretical investigations suggest this is possible~\cite{Chan:12}, experimentally trapping atoms, ions, or electrons on graphene surfaces remains more challenging than fabricating the nanotorus itself. Beyond the technical difficulties, Klein tunneling~\cite{Klein:29} has been identified as a fundamental obstacle to effectively confining relativistic electrons on graphene surfaces~\cite{Gutierrez:16,Renjun:18,Zhang:22}.

Once an electron is confined to the wormhole surface, $\mathcal{PT}$-phase transitions and Exceptional Points (EPs) can be explored in a manner analogous to trapped ion systems \cite{Li:2016tgr}. In general, the $\mathcal{PT}$ symmetry of a quantum system can be examined by analyzing its dynamics. Achieving $\mathcal{PT}$ symmetry fundamentally requires a precise balance between gain and loss, which presents significant challenges in quantum systems due to the inherent instability caused by gain-amplified noise \cite{doi:10.1126/science.aaw8205, Scheel_2018}. To mitigate this issue, passive systems exhibiting hidden $\mathcal{PT}$ symmetry have been proposed, where Hermitian systems are coupled to a dissipative reservoir \cite{PhysRevLett.103.093902, doi:10.1126/science.1258004}. The appearance of EPs in such lossy systems, despite the absence of active gain, has been experimentally demonstrated in optical and solid-state platforms \cite{Xiao}, providing a viable approach to observing and controlling quantum EP effects \cite{Xiao}. For instance, quantum coherence protection was observed in a $\mathcal{PT}$-broken superconducting circuit, although this required post-selection of results and exhibited an exponentially decreasing success rate over time. These groundbreaking studies on quantum EP systems \cite{doi:10.1126/science.aaw8205, Xiao} mark significant early steps toward the advancement of non-Hermitian quantum technologies.

\section{Conclusions\label{sec:Conclusions}}
In this work, we investigated non-Hermitian phase transitions with $\mathcal{PT}$ symmetry in a curved background described by a wormhole geometry. The analysis was carried out by studying the dynamics of a single fermion governed by the Dirac equation. Using the tetrad formalism and spin connection, we derived a second-order differential equation suitable for numerical analysis. This allowed us to compute the energy eigenvalues and explore the influence of the imaginary mass parameter $M=i\Gamma$ and the geometry of the surface, characterized by the radius $R$, on the structure of non-Hermitian phase transitions. We analyzed the behavior of the four lowest positive eigenvalues of $E^2$ as functions of $\Gamma$ and observed a sequence of well-defined non-Hermitian phase transitions characterized by multiple $\mathcal{PT}$-symmetric exceptional points. The lowest eigenvalue, $E_1$, exhibits an isolated exceptional point near the origin for small $\Gamma$. As $\Gamma$ increases, subsequent exceptional points emerge and are shared pairwise between adjacent eigenvalues: $E_1$ and $E_2$, $E_2$ and $E_3$, and $E_3$ and $E_4$. These exceptional points trace out imaginary-valued trajectories in the spectrum, indicating the nontrivial interplay between geometry and non-Hermitian dynamics. Furthermore, we studied the dependence of these exceptional points on the wormhole radius $R$, and found that the location of the exceptional points shifts with increasing $\Gamma$, while the sharing structure persists. Finally, we demonstrated that the deformation parameter $n$ also affects the position of the exceptional points. For large $n$, these points tend to cluster, a behavior reminiscent of the effects induced by the curvature parameter $\mathcal{R}$.

\section*{Acknowledgments}

This study was financed in part by the Coordenação de Aperfeiçoamento
de Pessoal de Nível Superior -- Brasil (CAPES) -- Finance Code 001.
YP would like to thank Isabella A. M. Brunetta for her hospitality
and support. J.F. thank the Fundação Cearense de Apoio ao Desenvolvimento
Científico e Tecnológico (FUNCAP) under the Grant No. PRONEM PNE0112-00085.01.00/16
for financial support, the CNPq under Grant No. 304485/2023-3, Gazi
University for the kind hospitality, Alexandra Elbakyan and Sci-Hub,
for removing all barriers in the way of science.

\bibliographystyle{apsrev4-2}
\bibliography{Reference}

\end{document}